\documentclass{svproc}
\usepackage{url}

\usepackage{amsmath}
\usepackage{mathtools}
\usepackage{commath}

\begin{document}
\mainmatter              
\title{Quick Lists: Enriched Playlist Embeddings for Future Playlist Recommendation}
\titlerunning{Quick Lists}  
%
\author{Brett Vintch}
\authorrunning{Brett Vintch} 
\institute{iHeartRadio\\ New York City, New York\\
\email{brett@iheart.com}}

\maketitle              

\begin{abstract}
Recommending playlists to users in the context of a digital music service is a difficult task because a playlist is often more than the mere sum of its parts. We present a novel method for generating playlist embeddings that are invariant to playlist length and sensitive to local and global track ordering. The embeddings also capture information about playlist sequencing, and are enriched with side information about the playlist user. We show that these embeddings are useful for generating next-best playlist recommendations, and that side information can be used for the cold start problem.
\end{abstract}

\section{Introduction}

Playlists are a common medium for music consumption and dissemination, and thus an important domain for the development of recommendation engines. While playlists are composed of individual tracks, the collection itself can be a distinct entity. Each track can be associated with multiple genres, moods, or concepts, and it is a track's context that defines its meaning and interpretation. For example, a Rage Against the Machine track could be included in both a rock genre playlist and in a protest song playlist, and the overall playlist context could plausibly affect a user's reaction to the track. In this work, we present a new method to embed playlists into a high dimensional space that is sensitive to local track context, and is naturally suited to recommending next-best playlists.

To fully capture the complexity of playlists, we believe that embeddings should meet a number of criteria. Embeddings should be invariant to playlist length and be sensitive to local or global track ordering. They should also ideally encode information about playlist sequencing, or the next-best future playlists given a current playlist. Much work has been done on embedding individual tracks using both user behavior \cite{pacula2009matrix,moore2012learning} and audio content \cite{van2013deep}, but it is not clear how one should aggregate these embeddings to the playlist level. Operations on individual item embeddings tend to employ order-invariant aggregations across the collection, such as sums, averages, or maximums. Though these approaches allow for comparison between playlists and are length-agnostic, they do not account for sequencing within a playlist or between playlists.

There are strong analogies between the task of representing playlists and that of representing natural language. Sentences are collections of words, where word order matters and phrase context gives additional meaning. Similarly, playlists are collections of tracks, where track ordering may be important and local track context can have an impact on the perceived mood or meaning. Recent works have made a similar connection \cite{mcfee2011natural}, but viable solutions mostly focus on recurrent neural networks for playlist completion \cite{keski2016music,vall2019order}. These types of recommenders are notoriously difficult to tune in order to produce useful recommendations, and they are also slow to train. Instead, we take inspiration from a new method that explicitly embeds sentences for use in determining the next most logical sentence  \cite{logeswaran2018efficient}; importantly, this method frames the process as a simple classification problem.

Our primary contribution is to show the utility of sentence embedding models for the task of recommending playlists. We extend this model to user side information and show that it is possible to manipulate recommendations with the addition of side information, and even to use only side information in a cold start situation. This new model meets the criteria for playlist embeddings outlined above, and is efficient to learn.

\begin{figure}[t!]
\begin{center}
\includegraphics[width=\linewidth]{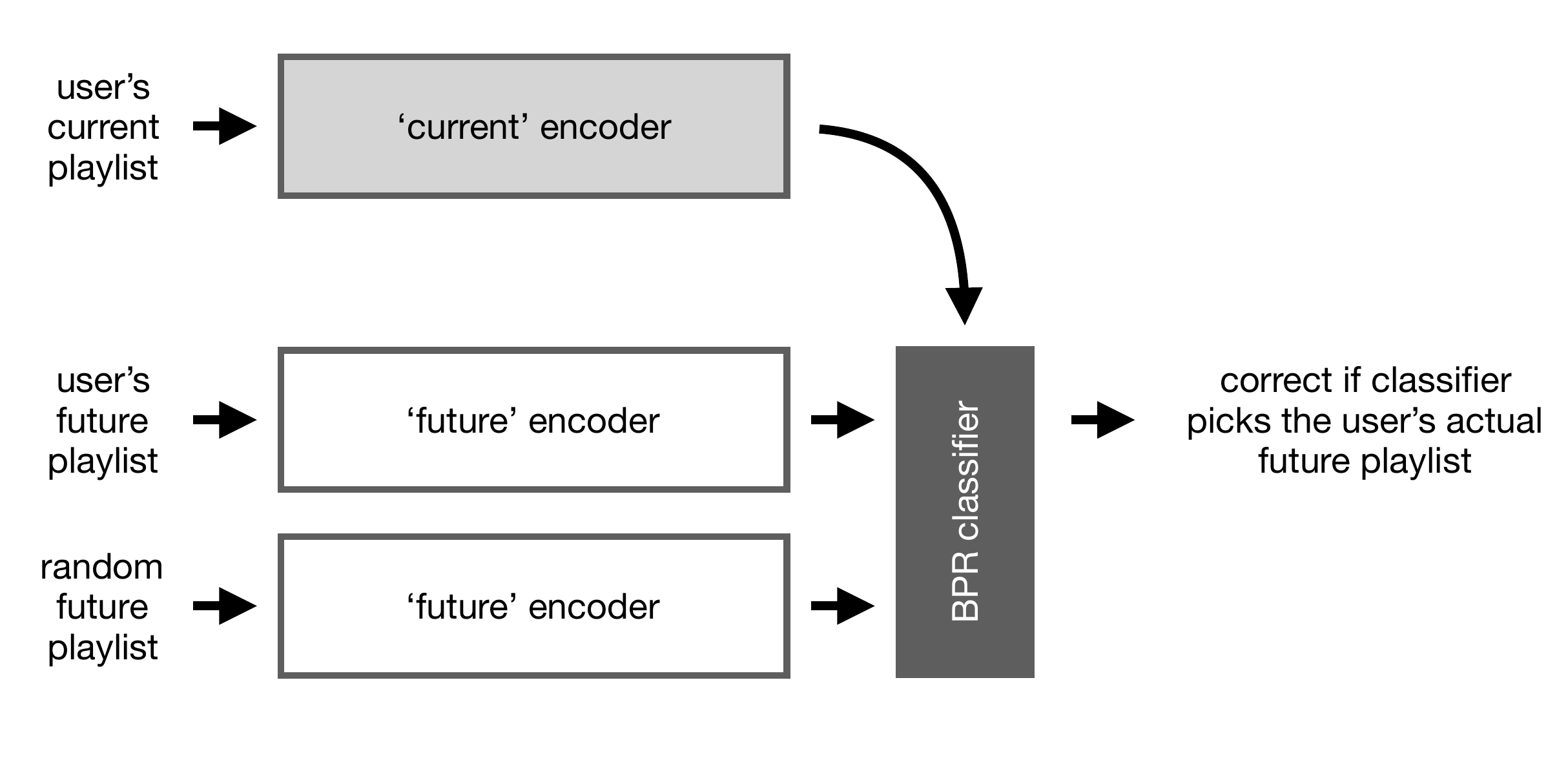} 
\end{center}
\caption[classification scheme]{Playlist recommendation is treated as a classification task. The training paradigm seeks to embed current and future playlists such that actual future playlists are selected with a higher probability than random playlists. The classifier is trained with a Bayesian Personalized Ranking (BPR) loss function}
\label{fig:classification}
\end{figure}

\section{Methods}

\subsection{Model}

The quick thoughts model, introduced by Logeswaran \& Lee \cite{logeswaran2018efficient}, treats sentence representation as a classification task. Sentences are encoded such that they are maximally predictive of the next sentence, as determined by a classifier. This discriminative approach to sentence embedding operates an order of magnitude faster than generative approaches, and learns to ignore aspects of sentences not connected to its meaning. Our approach for embedding playlists, quick lists, borrows this framework, substituting sequences of sentences with sequences of playlists. We further extend the framework by allowing for the inclusion of side information that describes the playlist listeners themselves.

\begin{figure*}[t!]
\begin{center}
\includegraphics[width=\textwidth]{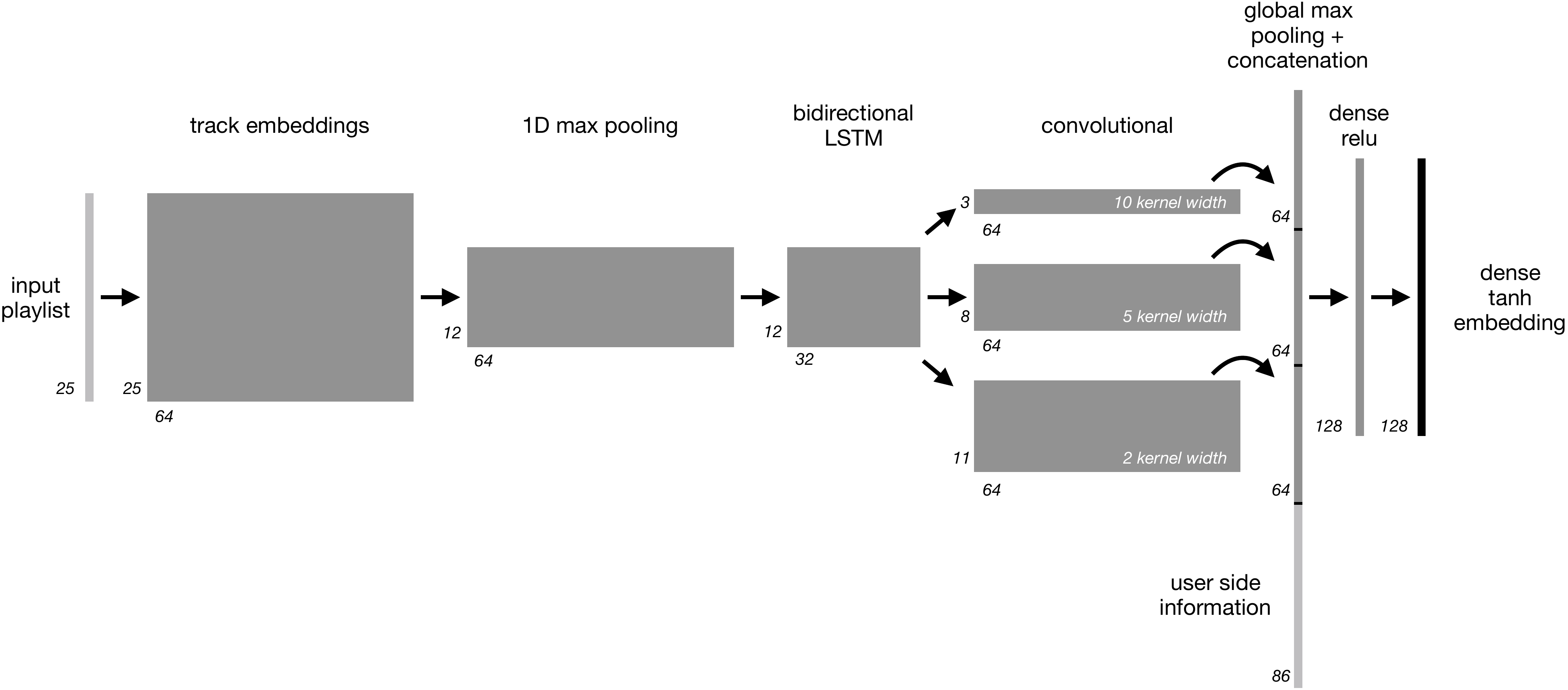} 
\end{center}
\caption[Embedding model]{Encoders embed playlists and user side information with a deep neural network that includes track embeddings, a bidirectional LSTM, a bank of convolutional kernels, and a final dense network. Numbers depict layer output sizes.}
\label{fig:network_diagram}
\end{figure*}

Our goal is to embed playlists such that embeddings are maximally predictive of the future playlists in a sequence. We define two encoders. The first, 'current' encoder embeds a playlist into a high dimensional space. The second, 'future' encoder embeds a user's subsequent playlist into the same space. A classifier then seeks to identify the correct playlist from a pair of playlists, where one is the actual future playlist and the other is a random playlist (Figure \ref{fig:classification}); that is, pairs of actual current and future playlists should be close together, and random playlists should be far apart. It is important that the current encoder and the future encoder are learned separately; although we want the embeddings from each encoder to live in the same space, current and future playlists are expected to be composed of different tracks that reflect a user's listening trajectory over time.

We chose our loss function to be analogous to Bayesian Personalized Ranking loss (BPR) \cite{rendle2009bpr}, which seeks to maximize the probability that a user $u$'s preferred item $p$ ranks higher than a user's non-preferred item $n$:

\begin{displaymath}
	P(p>n|\Theta,u) = \sigma(\hat{x}_{upn}(\Theta)),
\end{displaymath}
where $\sigma$ is a sigmoid function:

\begin{displaymath}
	\sigma(x) = \frac{1}{1 + e^{-x}}.
\end{displaymath}

$\hat{x}_{upn}(\Theta)$ is an arbitrary function parameterized by $\Theta$ that captures the relationship between a user's current and future playlists, $c$ and $p$, and compares it against the relationship between the user's current playlist and a random playlist, $n$. That is, $\hat{x}_{upn}(\Theta)$ captures the extent to which the actual future playlist is closer to the current playlist than a random future playlist.

We restrict our classifier to simple distance metrics so that learning is targeted to the playlist encoders and not the classifier; we prefer a powerful encoder for the generalizability of the embeddings. We considered euclidean distance, cosine distance, and a distance metric based on the dot product between two embeddings. Though dot products are commonly used for similarity calculations in recommendation tasks, we find that this metric's sensitivity to vector length encourages a bias towards popular content and does not produce qualitatively good predictions for less popular content. Though we observed that Euclidean-based models tended to take longer to converge, we also noticed a tendency for inference to be more equitable across content types; this was preferable for our use case, and so the experiments described below use euclidean distance. Thus,

\begin{displaymath}
	\hat{x}_{upn}(\Theta) = \norm{ \mathbf{v_{uc}}-\mathbf{v_{up}} } - \norm{ \mathbf{v_{uc}}-\mathbf{v_{n}} },
\end{displaymath}
where vectors $\mathbf{v_{ue}}$ represents playlist embeddings, and $\mathbf{e}$ can reference current ($c$), preferred ($p$), and non-preferred ($n$) playlists. 
The encoders for the current and next playlists both share the same architecture (Figure \ref{fig:network_diagram}), but they are trained independently so that they can adapt to sequence order. The encoders operate on a padded sequence of track vectors that are concatenated into a 2D matrix. This matrix is passed through a 1D max pooling function before being fed into a bidirectional LSTM with 16 hidden units. This output is then processed by a bank of 3 convolutional layers with different filter sizes (2, 5, and 10) and ReLu activation functions. Each filter output is subjected to another 1D max pooling function, and 50\% dropout is applied to this filter bank during training. The final output of the network is a dense layer with $Tanh$ activation functions and $L_2$ regularization; this layer produces the final mapping of each playlist to its embedding.

An additional factor often found in the context of playlists but not in natural language is the existence of user side information. We hypothesized that this information could be useful for recommendations, especially in the case of new users and cold starts. In the spirit of Wide and Deep models \cite{cheng2016wide}, we include a shallow network that combines categorical and scalar user information with the output of the encoder just before the final Tanh activation layer.

\subsection{Training}

We define and train the network in Keras \cite{chollet2015keras} with an Adam optimizer \cite{kingma2014adam}. Track embeddings are initialized as their word2vec embeddings learned over playlists as if they were sentences (we use the gensim implementation \cite{rehurek2010software} and drop tracks with 5 or fewer plays in our dataset). However, track embeddings are not fixed and are further learned along with the rest of the model during optimization. We find that fixing track embeddings hinders performance, and this seems to be especially true for euclidean-based classifiers.

The model is trained over 100 epochs using a learning schedule. The schedule drops the learning rate by a factor 0.25 every 10 epochs. Training takes about 16 hours on an NVIDIA Tesla K80. By manual inspection, epochs past the point where training loss begins to asymptote (10-20 epochs) help to fine tune the recommendations, and help most for users that play rare or unpopular content.

\subsection{Data}

The quick lists algorithm is designed to embed and recommend playlists. However, any ordered sequence of tracks can be used as input. Our primary use case is to recommend a next-best playlist to a user, and so for the purpose of these experiments we define a user's current playlist to be the sequence of most recently played tracks, regardless of their source. iHeartRadio has an array of digital music products, including live radio, custom artist radio, and user generated playlists. We take the last 50 played tracks for each user across all of these products and split them into 'current' and 'future' playlists. We do not include tracks that were thumbed down or skipped. In the case where a user listened to between 25 and 50 tracks, the last 25 tracks are assigned to the future playlist, and the rest are assigned to the current playlist.

Data is collected for a random sample of 300,000 users in January 2019. Users are further randomly split into training and testing sets with a 85/15 ratio. We also collect side information for each user where available. We one-hot encode the user's gender, age (in 5 year bins), and country of registration (out of 5 possible countries), and multi-hot encode their self-reported genre or radio format preferences from user on-boarding. There are 57 unique genres that were chosen, with the most popular being 'Top 40 \& Pop', 'Hip Hop and R\&B', and 'Country'. While a user's stated genre preference does not always reflect their revealed preference in actual listening, these preferences are of considerable interest to us as a possible solution to the cold start problem. In total, there are 86 binary features that describe a user and 72\% of users had at least one active feature.

\subsection{Experiments and analysis}

The quick lists procedure is intended to recommend future playlists to users based upon recent listening history. Playlist recommendation is treated as a nearest neighbor task. A user's current state, consisting of their recently played tracks and their profile side information, is embedded with the current encoder. Meanwhile, all possible future playlists are encoded with the future encoder. The future playlist that is closest to the current playlist in the embedded space is recommended to the user.

The embedded proximity of pairs of current and future playlists is an indicator of how well the model fits the data. During training we track the distribution of distances between current and future playlists and compare it to the distribution of distances between random current and future playlists. After fitting, we contrast these distributions to versions of the model that are effectively 'lesioned' by omitting the input data for different feature sets (set to zeros).

As a proxy for future playlist prediction accuracy we analyze the accuracy of track recommendations in predicted playlists. Specifically, we measure the overlap in tracks between the predicted playlists and each user's actual future playlist for the test set users, measured as an F1 score which combines precision and recall. We use this metric because it allows for easy comparisons between models, including baseline models. Note, however, that it does not take into account track order within playlists, despite sensitivity to order being a desired property; the metric may therefore miss more subtle quality differences between models. We also measure the percentage of tracks recommended in the predicted future playlist that also appear in a user's current playlist, as some use cases (including our own) may wish to penalize familiarity (i.e. repetitiveness) and encourage novelty.

We compare the quick lists model predictions to several baselines and to a word2vec-based approach. We consider a baseline where the recommended playlist is built from random tracks, one where the recommended playlist is an identical set of the most popular tracks in the data set, and one where the current playlist is simply repeated as the recommended future playlist. For the word2vec-based model, we average the word2vec vectors for each track in the current playlist and create a recommended playlist by finding the tracks closest to the average. For each of these approaches we draw the playlist length from the distribution of actual playlist lengths for test-set users.

\begin{figure}[t!]
\begin{center}
\includegraphics[width=\linewidth]{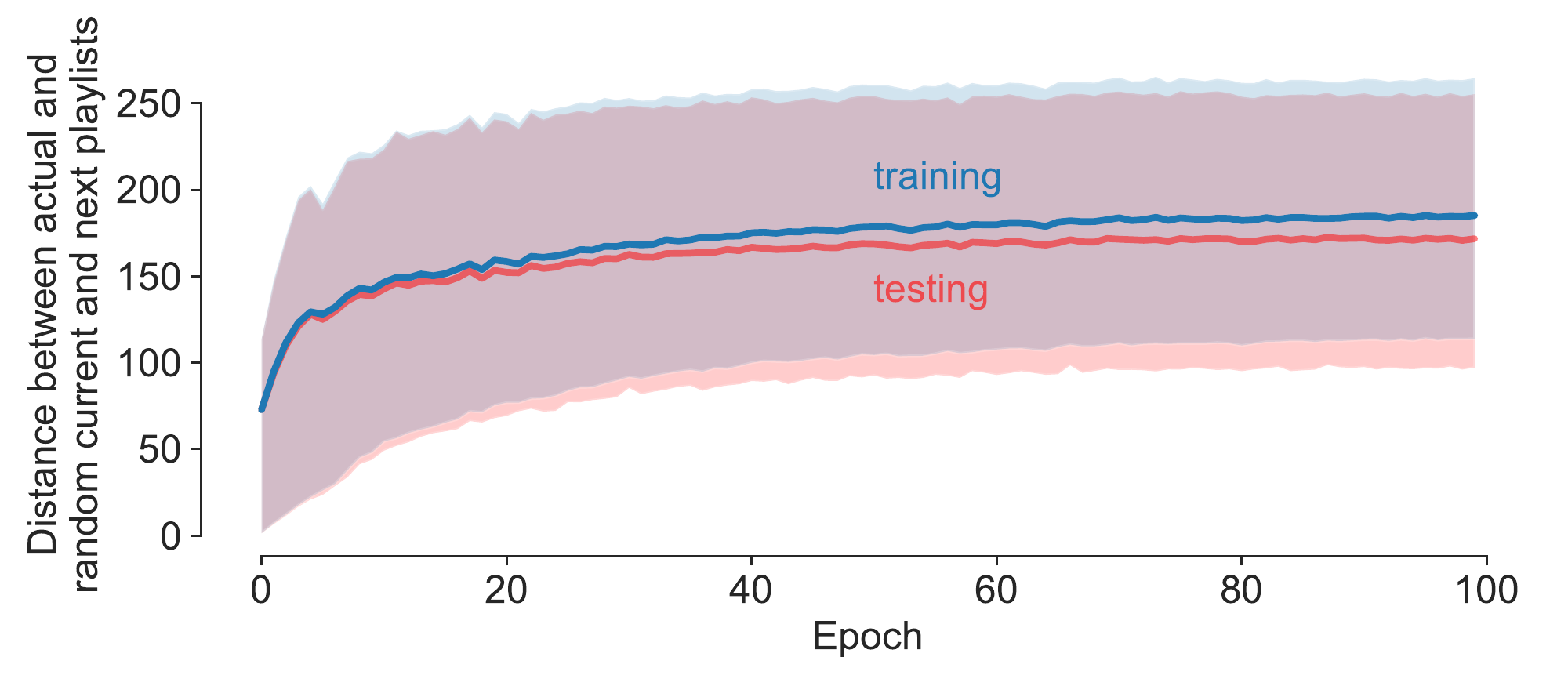} 
\end{center}
\caption[training distance]{Average (line) and 25\textsuperscript{th} and 75\textsuperscript{th} percentiles (band) of $\hat{x}_{upn}$ while training, for both training and test sets.}
\label{fig:training}
\end{figure}

\section{Results}

\subsection{Predictive}

\begin{figure}[t!]
\begin{center}
\includegraphics[width=\linewidth]{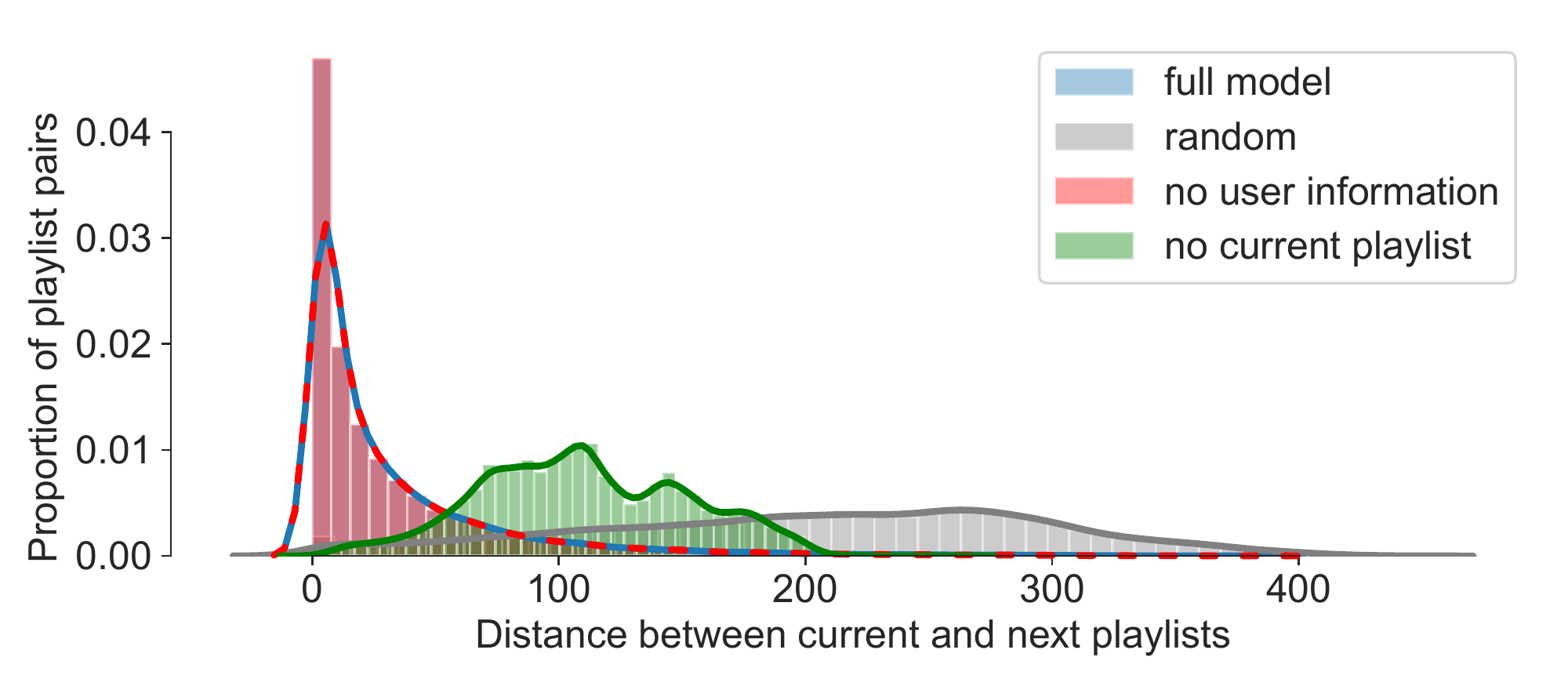} 
\end{center}
\caption[distribution]{Model performance measured as the distance between pairs of current and future playlists, with and without lesioning (red largely overlaps blue).}
\label{fig:lesioning}
\end{figure}

The quick list loss function encourages current and future user playlists to be close together in embedded space and random current and future playlists to be far apart. The model learns to distinguish these two categories during training with performance beginning to asymptote around 10-20 epochs, for both training and testing data (Figure \ref{fig:training}). It continues to improve in subsequent epochs but at a slower rate. We justify the continued training by observing qualitative improvements for less popular content.

We assess the relative importance of each of the two model inputs by omitting one at a time during inference. We use the distance between actual pairs of current and future playlists as a measure of performance quality for users in the test set, where the desired distance of zero would so that real current and future playlist embeddings perfectly overlap. The distribution of distances across users for the full model and the model without side information show similar performance (Figure \ref{fig:lesioning}; blue and red distributions, which largely overlap), with the lesioned model performing only 0.5\% worse than the full model on average. This is an indication that side information is less informative in predicting future playlists than a user's recent listening history. Alternatively, removing recent playlist information reduces average performance by 346\% (green). Reversing the current playlist before embedding also leads a decrease in performance of 18.5\% on average (not shown), which indicates the importance of track ordering in deriving embeddings. However, all models perform better than a weak baseline, where current and future playlists are paired randomly (gray; 645\% decrease in performance compared to the full model, on average). Thus, even the scenario where side information is used alone shows some some predictive power.

\begin{table*}[h!]
  \begin{center}
    \caption{Model performance}
    \label{tab:performance}
    \begin{tabular}{l | r | r} 
      Model & F1 & Familiarity \\
      \hline
      Baseline - Random tracks & 0.00022 & 0.014\% \\
      Baseline - Popular tracks & 0.026 & 2.1\% \\
      Baseline - Current playlist as future playlist & \textbf{0.092} & 100\% \\
      Word2vec - Closest tracks to average & 0.020 & 2.8\% \\
      \hline
      Quick lists - No current playlist & 0.018 & 1.5\% \\
      Quick lists - No user information & \textbf{0.076} & 7.5\% \\
      Quick lists - Full model & 0.075 & 7.5\%  \\
      Quick lists - Reversed current playlist order & 0.072 & 7.5\% \\
       \end{tabular}
  \end{center}
\end{table*}

We also examine recommendation quality by measuring the frequency by which tracks in the recommended playlist actually occur in the future playlist. We measure this overlap as the F1 score between the predicted future playlist and the actual future playlist for each user in the test set. We also measure the percentage of tracks in the predicted future playlist that are found in the current playlist as a measure of familiarity (low overlap means lower familiarity but higher novelty, which is generally preferable). Table \ref{tab:performance} shows these metrics for a collection of baseline models and quick list models. The quick lists model performs relatively well compared to most baseline models, with only moderate repetition between the recommended future playlist and the current playlist. Reversing the current playlist order reduced predictive power slightly, but removing information about the current playlist drastically decreases accuracy. This lesioned model, however, does still have some predictive power above random playlists, and may still be useful for cold start users (see the Qualitative section below). In the context of the test set users, the lesioned model with no user information slightly outperforms the full model.

Among the baseline models, simply using the current playlist as the recommended future playlist performs surprisingly well, beating the best quick lists model F1 score. With a repetition rate of 100\% the recommended playlist is a poor user experience and thus not viable for most production purposes. However, it does demonstrate real users preferences for familiarity and repetition.

\subsection{Qualitative}

\begin{figure*}[t!]
\begin{center}
\includegraphics[width=\textwidth]{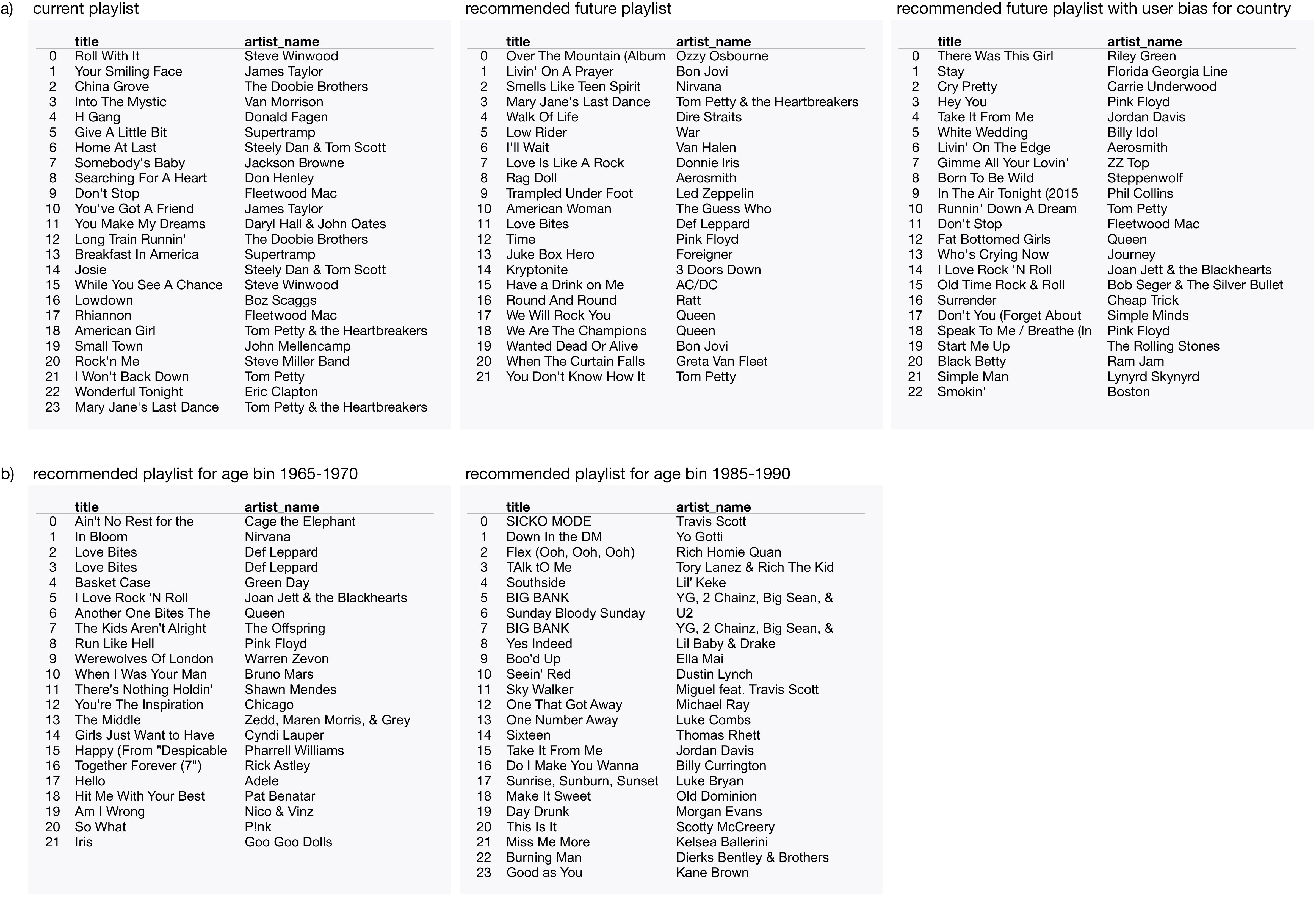} 
\end{center}
\caption[manipulated_playlistsl]{Examples of manipulating a user's side information to generate playlists. a) Recommended playlists with manipulation of side information for an actual user. Left: actual current playlist. Middle: recommended future playlist for this user. Right: recommended future playlist with an artificial preference for Country music injected via side information. b) Recommended playlists for a new user with no current playlist and only one active age bin feature.}
\label{fig:manipulated_playlists}
\end{figure*}

Recommendations can also be generated for manipulated or arbitrary inputs. For example, if we start with a user that most recently listened to songs in the classic rock genre, but inject a strong preference for country music  via the user's side information, we can change the user's recommended future playlist from pure rock to something that crosses the boundaries between classic rock and country (Figure \ref{fig:manipulated_playlists}a).  Similarly, we can use a user's side information to help solve the cold start problem. Recommended playlists can be generated for hypothetical new users where the model is only supplied with the user's age range (Figure \ref{fig:manipulated_playlists}b).

Finally, we observe that track order is important in generating embeddings for use in recommending future playlists. We take two playlists that contain the same set of ten tracks; in one, tracks are ordered in a progression from the alternative tracks to the classic rock tracks, and the other playlist they are reversed. Despite an identical set of tracks, each input produces a recommendation that emphasizes continuity with the tracks that were most recently played. The first two recommended tracks for the current playlist ending with alternative tracks are from the artists Flora Cash and Fall Out Boy, while they are from the artists Supertramp and Bonnie Tyler for the current playlist ending in classic rock (full playlists not shown for space). 

\section{Discussion}

We present a novel method to embed playlists and use those embeddings to make recommendations for future playlists. This method builds upon recent advances in natural language representation for embedding sentences, and adds the ability to leverage side information for the playlist user. Though side information alone does not appear to provide very accurate recommendations compared to recent listening history, we demonstrate that it may still be useful for the cold start problem and for playlist manipulation.

Real listeners demonstrate repetitive behavior, listening to a handful of tracks many times. This pattern leads to the surprising result that simply using a user's current playlist as a prediction for their best future playlist is reasonably accurate approach. Prior work has indeed shown a reliable preference of real listeners for familiar music \cite{ward2014same}. Unfortunately, for real world music recommendation products this simple tactic is usually not a viable solution because the user experience is undesirable.

In the experiments described above we define a playlist as a collection of tracks listened in sequence, regardless of their source. This liberal definition was chosen because in our use case we wish to make future playlist recommendations to users using their most recent history, regardless of how they listened to each track. However, this definition may also increase problem difficulty as compared to a scenario in which a user listened to user-generated or curated playlists. This is because these types of playlists are more likely to be narrow in scope and coherent in theme. Despite this added difficulty, we find that the model trained with the more liberal definition of a playlist still produces useful recommendations.

A logical next step in this work is to improve the decoder for recommendation. In this work, we rely on recommending playlists that other users have actually created through the use of nearest neighbor lookup. However, there is nothing barring one from training a separate decoder that takes playlist embeddings and produces new, never before created playlists. We have begun experimenting with the use of recurrent neural networks, specifically LSTMs, to create a generator for playlists that are potentially infinitely long; we see encouraging results thus far.

\bibliographystyle{IEEEtran}

\bibliography{bibliography}

\end{document}